\documentclass[onecolumn,amsmath,amssymb,nofootinbib,12pt]{article}
\usepackage{jheppub}
\usepackage{ifpdf}
\usepackage{graphicx,subcaption}
\usepackage{amsfonts}
\usepackage{amsmath}
\usepackage{amssymb}
\usepackage{epsfig,bm}
\usepackage{pdfpages}
\usepackage{graphicx,epstopdf}
\usepackage[makeroom]{cancel}
\hypersetup{pdftex,colorlinks=true,allcolors=blue}
\usepackage{array}
\usepackage{ulem}

\def\be{\begin{equation}}
\def\ee{\end{equation}}
\def\ba{\begin{eqnarray}}
\def\ea{\end{eqnarray}}

\def\nn{\nonumber}
\def\lf{\left}
\def\rt{\right}

\newcommand{\eq}[1]{(\ref{#1})}

\makeatletter

\newcommand{\Rmnum}[1]{\expandafter\@slowromancap\romannumeral #1@}
\makeatother
\def\nn{\nonumber}\def\lf{\left}\def\rt{\right}\def\q{\theta}   \def\y {\psi}   \def\p {\pi} \def\a {\alpha}  \def\d {\delta} \def\f {\phi} \def\g {\gamma} \def\h {\eta}  \def\k {\kappa} \def\l {\lambda} \def\z {\zeta} \def\x {\xi} \def\c {\chi}  \def\n {\nu} \def\m {\mu} \def\pd {\partial} \def \inf {\infty}  
\def\Q{\Theta}  \def\Y {\Psi}    \def\S {\Sigma}  \def\F {\Phi}      \def\grad{\nabla}\def\.{\cdot}
\def\math {\mathcal}
\usepackage{xspace}
\begin{document}
\title{\Large Holographic complexity of the electromagnetic black hole}
\author[a]{Jie Jiang, }
\affiliation[a]{Department of Physics, Beijing Normal University,
Beijing 100875, China}
\author[a,b]{Ming Zhang}
\affiliation[b]{College of Physics and Communication Electronics, Jiangxi Normal University, Nanchang 330022, China}
\emailAdd{jiejiang@mail.bnu.edu.cn, mingzhang@mail.bnu.edu.cn}
\date{\today}
\abstract{
In this paper, we use the ``complexity equals action" (CA) conjecture to evaluate the holographic complexity in some multiple-horzion black holes for F(Riemann) gravity coupled to a first-order source-free electrodynamics. Motivated by the vanishing result of the purely magnetic black hole founded by Goto $et.\, al$, we investigate the complexity in a static charged black hole with source-free electrodynamics and find that this vanishing feature of the late-time rate is universal for a purely static magnetic black hole. However, this result shows some unexpected features of the late-time growth rate. We show how the inclusion of a boundary term for the first-order electromagnetic field to the total action can make the holographic complexity be well-defined and obtain a general expression of the late-time complexity growth rate with these boundary terms. We apply our late-time result to some explicit cases and show how to choose the proportional constant of these additional boundary terms to make the complexity be well-defined in the zero-charge limit. For the static magnetic black hole in Einstein gravity coupled to a first-order electrodynamics, we find that there is a general relationship between the proper proportional constant and the Lagrangian function $h(\math{F})$ of the electromagnetic field: if $h(\math{F})$ is a convergent function, the choice of the proportional constant is independent on explicit expressions of $h(\math{F})$ and it should be chosen as $4/3$; if $h(\math{F})$ is a divergent function, the proportional constant is dependent on the asymptotic index of the Lagrangian function.
}
\maketitle
\section{Introduction}
In recent years, there has been a growing interest in the topic of ``quantum complexity", which is defined as the minimum number of gates required to obtain a target state starting from a reference state \cite{L.Susskind,2}.  From the holographic viewpoint, Brown $et\,al.$ suggested that the quantum complexity of the state in the boundary theory is dual to some bulk gravitational quantities
which are called ``holographic complexity". Then, the two conjectures, ``complexity equals volume" (CV) \cite{L.Susskind,D.Stanford} and ``complexity equals action" (CA) \cite{BrL,BrD}, were proposed. They aroused researchers' widespread attention to both holographic complexity and circuit complexity in quantum field theory, $e.g.$ \cite{Jiang:2019pgc,Bhattacharyya:2018bbv,Ali:2018fcz,Ali:2018aon,HosseiniMansoori:2018gdu,Liu:2019smx,
A51,A11,A12,A13,A14,A15,A16,A17,A18,A19,A20,A23,A24,A25,A26,A27,A28,A29,A30,A31,A32,A33,A34,A35,A36,
Fan:2019mbp,Goto:2018iay,JiangL,Nally:2019rnw,Jiang3,Chapman1,Chapman2,Jiang4,SZ,Roberts:2014isa,A37,A38,A39,Jiang1,A2,A3,A4,A5,A6,A7,A8,A9,A10}.

In present work, we only focus on the CA conjecture, which states that the quantum complexity of a particular state $|\y(t_L,t_R)\rangle$ on the boundary is given by
\ba\label{CA}
C\lf(|\y(t_L,t_R)\rangle\rt)\equiv\frac{I_\text{WDW}}{\p\hbar}\,.
\ea
Here $I_\text{WDW}$ is the on-shell action in the corresponding Wheeler-DeWitt (WDW) patch,
which is enclosed by the past and future light sheets sent into the bulk spacetime from the timeslices $t_L$ and $t_R$.

By studying a simple class of systems known as random quantum circuits with $N$ qubits, it has been shown that for generic circuits, after a short period of transient initial behavior, the complexity grows linearly in time, and finally saturates at a maximum value. In the context of AdS/CFT, we need to set $N$ to be very large. Then, it can be generally argued that at late times, this quantum complexity should continue to grow with a rate given by\cite{L.Susskind,D.Stanford}
\ba\label{TS}
\frac{dC}{dt}\sim TS\,,
\ea
where the entropy represents the width of the circuit and the temperature is an obvious choice for the local rate at which a particular qubit interacts.

Recently, Goto $et.\,al$\cite{Goto:2018iay} investigated the CA complexity for $dyonic$ Reissner-Nordstrom-AdS(RN-AdS) black holes in $4$-dimensional Einstein-Maxwell gravity. They found a surprising result that the growth rate vanishes at late times when this $dyonic$ black hole only carries a magnetic charge. However, this result does not agree with the general expectation \eq{TS} for the quantum system. Moreover, from the perspective of the boundary CFT, nothing particularly strange should happen in the zero-charge limit. Therefore, the holographic complexity should also satisfy this limit. But this result also shows the unexpected feature in the zero-charge limit, i.e., the limit of complexity for the charged black hole should be as same as the neutral counterpart.

However, this apparent failure can be alleviated when we modify the total action with the addition of the Maxwell
boundary term\cite{Goto:2018iay}
\ba
I_{\m\text{Q}}=\frac{\g}{4\p}\int_{\pd M}d\S_aF^{ab}A_b
\ea
for the Einstein-Maxwell gravity. Here $\g$ is some proportional constant which should be chosen as $\g=1$ for the purely magnetic $dyonic$ black hole to ensure the complexity satisfies the zero-charge limit. After that, the late-time rate becomes finite and sensitive to the magnetic charge. Moreover, we can see that this boundary term does not affect the equation of motion of the electromagnetic fields. It only changes the boundary conditions in the variational principle of the electrodynamics.

To better understand the vanishing of the late-time complexity growth rate, we might also ask whether this result is universal in a purely magnetic black hole with a source-free electromagnetic field. Therefore, in this paper, we would like to investigate the holographic complexity of a magnetic black hole for a gravitational theory coupled to source-free electrodynamics and try to discuss what appropriate boundary action can be introduced to make the holographic complexity be well-defined.

The remainder of this paper is organized as follows: in Sec.\ref{Sec2}, we review the Iyer-Wald formalism for an invariant gravity coupled to a first order source-free electrodynamics. In Sec.\ref{Sec3}, we evaluate the late-time holographic complexity growth rate for the original CA conjecture as well as the new conjecture with some additional boundary terms in a static multiple-horizon black hole for F(Riemann) gravity coupled to source-free fields. In Sec.\ref{Sec4} and Sec.\ref{Sec5}, we apply our late-time result to the $dyonic$ black hole in Maxwell-$f(R)$ gravity and charged dilaton black hole, individually. In Sec.\ref{Sec6}, we apply our result to some purely magnetic black holes in Einstein gravity. First, we investigate some static magnetic black holes in Einstein gravity coupled a electromagnetic field with some special Lagrangian functions in Sec.\ref{Sec61} and \ref{Sec62}, and show how to fix the proportional constant to make the complexity be well defined in these explicit cases. Then, in Sec.\ref{Sec63}, we give a general discussion of the static magnetic black hole in the Einstein gravity with the first order electrodynamics. Finally, concluding remarks are given in Sec.\ref{Sec6}

\section{Iyer-Wald formalism}\label{Sec2}
In this section, we will give a brief review of the Iyer-Wald formalism for a general $4$-dimensional diffeomorphism invariant theory coupling a first-order electromagnetic field and source-free scalar field, which is described by a Lagrangian $\bm{L}=\math{L}\bm{\epsilon}$ where the dynamical field consists of a Lorentz signature metric $g_{ab}$, gauge field $A_a$ and a scalar field $\y$. Following the notation in \cite{IW}, we use boldface letters to denote differential forms and collectively refer to $(g_{ab},A_a,\y)$ as $\f$. Then, the action can be divided into the gravity part, gauge field part and scalar field part, i.e., $\bm{L}=\bm{L}_\text{grav}-\bm{L}_\text{em}+\bm{L}_\y$
where  $\bm{L}_\text{em}=\math{L}_\text{em}\bm{\epsilon}=h(\bm{F},\y)\bm{\epsilon}$ and $\bm{L}_\y=\math{L}(\y,|\grad \y|^2)\bm{\epsilon} $.
Here $\bm{F}=d\bm{A}$ is the electromagnetic tensor and $|\grad\y|^2=\grad_a\y\grad^a\y$. The variation of the gravitational
part with respect to $g_{ab}$ is given by
\ba
\d \bm{L}_\text{grav}=\bm{E}_{g}^{ab}(\f)\d g_{ab}+d \bm{\Q}(\f,\d g)\,,
\ea
where $\bm{E}_g^{ab}(\f)$ is locally constructed out of $\f$ and its derivatives and $\bm{\Q}$ is locally constructed out of
$\f,\, \d g_{ab}$ and their derivatives. The equation of motion can be read off as
\ba
\bm{E}_g^{ab}(\f)=-\frac{1}{2}T^{ab}\bm{\epsilon}\,
\ea
with
\ba\label{Tab}
T^{ab}=-\frac{2}{\sqrt{-g}}\frac{\d \sqrt{-g}\math{L}_\text{mt}}{\d g_{ab}}=-g^{ab}\math{L}_\text{mt}-2\frac{\d \math{L}_\text{mt}}{\d g_{ab}}\,,
\ea
which is the stress-energy tensor of the matter fields. Here we denote $\bm{L}_\text{mat}=\math{L}_\text{mt}\bm{\epsilon}=\lf(\math{L}_\y-\math{L}_\text{em}\rt)\bm{\epsilon}$. Let $\z^a$ be the infinitesimal generator of a diffeomorphism. Exploiting the Bianchi identity $\grad_{a}T^{ab}=0$, one can obtain the identically conserved current for a generic background metric $g_{ab}$ as
\ba\label{J1}\begin{aligned}
\bm{J}[\z]&=\bm{\Q}(\f,\z)-\z\.\bm{L}_\text{grav}
+s_{\z}\.\bm{\epsilon}\,,
\end{aligned}\ea
where $s^a_{\z}\equiv -T^{ab}\z_b$ and  $\bm{\Q}(\f,\z)=\bm{\Q}(\f,\math{L}_{\z}g_{ab})$. Since $\bm{J}$ is closed, there exists
a Noether charge $2$-form $\bm{K}[\z]$ such that $\bm{J}[\z]=d\bm{K}[\z]$. With similar arguments in \cite{IW,WSS}, this
$2$-form can always be expressed as
\ba\label{K}
\bm{K}[\z]=\bm{W}_c\z^c+\bm{X}^{cd}\grad_{[c}\z_{d]}\,,
\ea
where
\ba\label{Xcd}
\lf(\bm{X}^{cd}\rt)_{c_1c_2}=-E_R^{abcd}\bm{\epsilon}_{abc_1c_2}\,
\ea
is the Wald entropy density with
\ba
E_R^{abcd}=\frac{\pd \math{L}_\text{grav}}{\pd R_{abcd}}\,.
\ea
Substituting (\ref{Tab}) into (\ref{J1}), one can obtain
\ba\label{Lz}
\z\.\bm{L}=\bm{\Q}(\f,\z)-d \bm{K}[\z]+\c_\z\.\bm{\epsilon}\,,
\ea
where we denote
\ba\begin{aligned}\label{ca}
\c^a_\z=-2\frac{\d \math{L}_\text{mt}}{\d g_{ab}}\z_b\,.
\end{aligned}\ea
Then, we consider the electromagnetic part $\bm{L}_\text{em}$.  Since $h(\bm{F},\y)$ and $\y$ are scalar fields, all of the indexes should be contracted. Then, the Lagrangian can be expressed as a function of the scalar fields
\ba
\math{F}^{(n)}=F_{a_1}{}^{a_2}F_{a_2}{}^{a_3}\cdots F_{a_{n-1}}{}^{a_n}F_{a_{n}}{}^{a_1}\,,
\ea
i.e., $\bm{L}_\text{em}=h\lf(\math{F},\y\rt)$ with $\math{F}=\{\math{F}^{(2)}, \math{F}^{(4)}, \cdots \math{F}^{(2n)}, \cdots \}$. For latter convenience, here we also define a tensor
\ba
H^{(n)}_{ab}=F_{a}{}^{a_2}F_{a_2}{}^{a_3}\cdots F_{a_{n-1}}{}^{a_n}F_{a_n b}\,.
\ea
With these in mind, the variation of the electromagnetic part with respect to $\bm{A}$ is given by
\ba\begin{aligned}
\d \bm{L}_\text{em}&=\sum_{n}h_n\d(\star\math{F}^{(2n)})\\
&=2\sum_{n}n h_n\star\lf(H^{(2n-1)}_{ab}\d F^{ba}\rt)\\
&=-4\lf[\sum_{n}n h_n \star\bm{H}^{(2n-1)}\rt]\wedge\d\bm{F}\\
&=\bm{G}\wedge d\d\bm{A}\\
&=-d\bm{G}\wedge \d\bm{A}+d\lf(\bm{G}\wedge \d\bm{A}\rt)\,.
\end{aligned}\ea
where we have denoted
\ba
h_n=\frac{\pd h(\math{F},\y)}{\pd \math{F}^{(2n)}}
\ea
and define $\bm{G}=\star \bm{H}$ with
\ba\label{Hab}
\bm{H}=-4\lf[\sum_{n}n h_n \bm{H}^{(2n-1)}\rt]\,.
\ea
We have also used the relation
\ba
F_{1ab}F_2^{ab}=-2\star\lf(\bm{F}_1\wedge\star \bm{F}_2\rt)
\ea
for two $2$-form $\bm{F}_1$ and $\bm{F}_2$. Since the scalar field is source-free, the equation of motion for the electromagnetic field is given by $d\bm{G}=0$, which is also
equivalent to
\ba
\grad_a\lf(H^{ab}\rt)=0\,.
\ea

Next, we turn to evaluate \eq{ca}. If $\z$ is a Killing vector, i.e., $\math{L}_\z\bm{A}=0$ and $\math{L}_\z \y=0$, we have
\ba\begin{aligned}
&\c^a_\z=-2\frac{\d h(\math{F},\y)}{\d g_{ab}}\z_b-2\frac{\d \math{L}(\y,|\grad \y|^2)}{\d g_{ab}}\z_b\\
&=-4\sum_{n}n h_n \lf(H^{(2n-1)}\rt)^{ac}F_c{}^b\z_b-2\frac{\pd \math{L}}{\pd |\grad \y|^2}\grad^a \y\math{L}_\z \y\\
&=H^{ac}F_c{}^b\z_b\\
&=\grad_b\lf(H^{ba}A_c\z^c\rt)\,,
\end{aligned}\ea
which implies
\ba
\c_\z\cdot\bm{\epsilon}=d \lf(A_a\z^a\bm{G}\rt)\,.
\ea
Combing with the fact $\bm{\Q}(\f,\z)=0$ for the Killing vector, Eq. \eq{Lz} becomes
\ba\label{kL}
\z\cdot \bm{L}=d\lf(A_a\z^a\bm{G}-\bm{K}[\z]\rt)\,.
\ea
Moreover, the equation of motion $d\bm{G}=0$ implies that $\bm{G}$ is a closed form for the on-shell field. Then, there exists
a $1$-form $\bm{B}$ such that $\bm{G}=d\bm{B}$ when the EM field satisfies the equation of motion. Combing the Bianchi identity
$d\bm{F}=0$, the electric charge $Q$ and magnetic charge $P$ can be defined as
\ba\begin{aligned}
Q=\int_{\math{C}_\inf}\bm{G}\,,\ \ \ \ P=\int_{\math{C}_\inf}\bm{F}\,,
\end{aligned}\ea
where $\math{C}_\inf$ denotes a $2$-dimensional surface at the asymptotic infinity.
\begin{figure}
\centering
\includegraphics[width=0.6\textwidth]{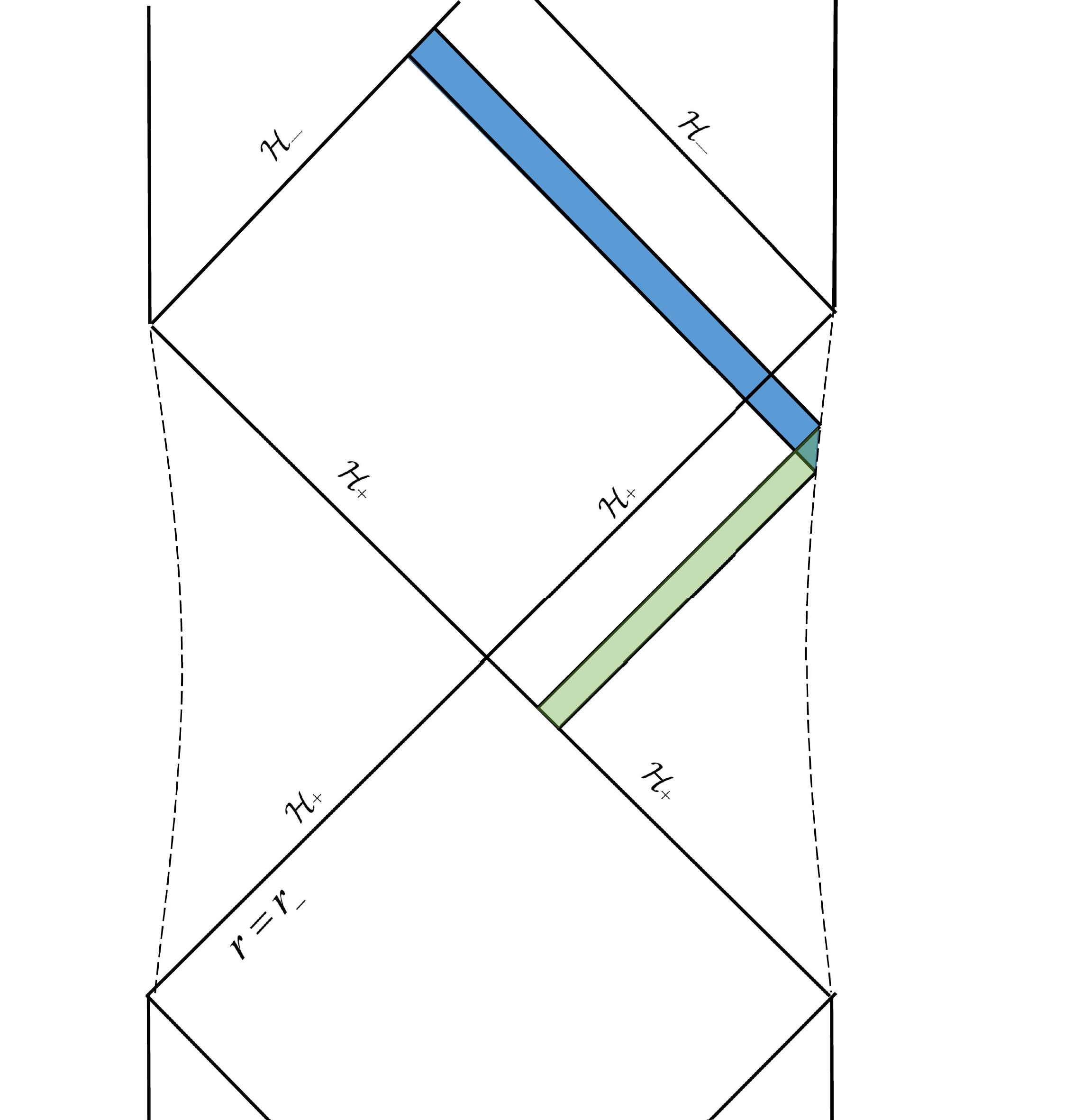}
\caption{Wheeler-DeWitt patch at late time of a multiple Killing horizon black hole, where the dashed lines denote the cut-off surface at asymptotic infinity, satisfying the asymptotic symmetries. }\label{WDW}
\end{figure}
\section{Late-time complexity growth rate}\label{Sec3}
\subsection{Original CA conjecture}\label{Sec31}
In this subsection, we consider a static magnetic black hole with the Killing horizon contained a bifurcation surface. And $\x^a=(\pd/\pd t)^a$ is the Killing vector of this horizon. By using this static Killing vector, we can define the electric potential and magnetic potential
\ba
\F=-A_a\x^a\,,\ \ \ \Y=B_a\x^a\,.
\ea

According to the CA conjecture, calculating the holographic complexity is equivalent to evaluating the full action within the WDW patch. For a $F($Riemanm$)$ gravity, the full action can be expressed as\cite{Jiang3}
\ba\label{fa}\begin{aligned}
I=&\int_{M} \bm{L}+\int_{\math{C}}\bm{s}\h
+\int_{\math{N}} d\l\bm{s}\k+\int_{\math{N}} d\l (\pd_\l\bm{s})\log\lf(l_\text{ct}\Q\rt),\nn
\end{aligned}\\\ea
where $\bm{s}=\bm{X}^{cd}\bm{\epsilon}_{cd}$ is the Wald entropy density, $\l$ is the parameter of the null generator $k^a$ on the null segment, $\k$ measures the failure of $\l$ to be an affine parameter which is derived from $k^a\grad_ak^b=\k k^b$, $\Q=\nabla_ak^a$ is the expansion scalar, and $l_\text{ct}$ is an arbitrary length scale. These boundary terms are introduced to make the variational principle well-posed.

First of all, we consider the bulk contributions of the change of the action. As illustrated in Fig.1, the bulk contributions only come from the region $\d M$. At the late times, it can be generated by the Killing vector $\x^a$ through the null boundary $\math{N}$ which is bounded by the inner and outer horizons. Then, the action change contributed by the late-time bulk action can be shown as
\ba
\d I_\text{bulk}=I_{\d M_2}-I_{\d M_1}\,,
\ea
For simplification, we will neglect the index $\{\pm\}$. Turning to the bulk contribution from $\d M$, we have
\ba\begin{aligned}
I_{\d M}&=\int_{\d M}\bm{L}=\d t\int_{\math{N}}\x\.\bm{L}\,.\\
\end{aligned}\ea
Using \eq{kL}, we have
\ba\begin{aligned}\label{xL}
\int_{\math{N}}\x\.\bm{L}&=-\int_{\math{N}}d\lf(\F\bm{G}+\bm{K}[\x]\rt)\\
&=-\int_{\math{C}_\inf}\F\bm{G}-\int_{\math{C}_\inf}\bm{K}[\x]
+\int_{\math{C}}\F\bm{G}+\int_{\math{C}}\bm{K}[\x]\,.\\
\end{aligned}\ea
At the late times, the corner $\math{C}$ approaches the Killing horizon $\math{H}$. Since the horizon contains a bifurcation surface,
the first term in \eq{K} vanishes on the horizon $\math{H}$, i.e.,
\ba
\bm{K}=\bm{X}^{cd}\grad_{[c}\x_{d]}=\k\bm{s}\,,
\ea
where $\bm{\epsilon}_{ab}$ is the binormal of surface $\math{C}$, and $\x^a\grad_a\x^b=\k\x^b$ on the horizon. By virtue of the smoothness of the pullback of $A_a$ and the static condition, one can show that $\F_{\math{H}}=-\left.\x^aA_a\right|_{\math{H}}$ is constant in the portion of the horizon to the future of the bifurcation surface. For simplification, we can choose $\left.A_a\x^a\right|_{\inf}=0$. With these in mind, \eq{xL} becomes
\ba
\int_{\math{N}}\x\.\bm{L}=T S+\F_\math{H} Q-\int_{\math{C}_\inf} \bm{K}[t]
\ea
with the entropy $S=2\p\int_{\S}\bm{X}^{cd}\bm{\epsilon_{cd}}$ and $T=\k /2\p$. With these in mind, we can obtain
\ba\label{dIdt}
\frac{d I}{dt}=\left[TS+\F_\math{H}Q\right]^{-}_{+}\,,
\ea
where the index $\pm$ present the quantities evaluated at the ``outer" or first ``inner" horizon.

Next, we consider the boundary and corner contributions to action growth. Without loss of generality, we shall adopt the affine parameter for the null generator of the null surface. As a consequence, the surface term vanishes on all null boundaries. Meanwhile, we choose $l^a$ as the null generator of the null boundary $\math{N}$, in which $l^a$ satisfies $\math{L}_{\x}l^a=0$. Then, the time derivative of the counterterm contributed by $\math{N}$ vanishes. By considering that the entropy is a constant on the Killing horizon, i.e., $\math{L}_{\x}\bm{s}=0$, the counterterm contributed by the null segment on the horizon also vanishes.

The affinely null generator on the horizon can be constructed as $k^a=e^{-\k\l}\x^a=e^{-\k\l}\lf(\frac{\pd}{\pd \l}\rt)^a$. The
transformation parameter can be shown as\cite{Jiang3}
\ba
\h(\l)=\ln\lf(-\frac{1}{2}k\.l\rt)=-\k\l+\ln\lf(-\frac{1}{2}\x\.l\rt)\,.
\ea
Then, we have
\ba
\frac{d I_\text{corner}}{d t}=\frac{d I_\text{corner}}{d \l}=-T S\,.
\ea
Combining those contributions, we have
\ba\label{dIdt}
\lim_{t\to\inf}\frac{d C_A}{dt}=\frac{1}{\p \hbar}\lf(\F_{\math{H}_-}Q-\F_{\math{H}_+}Q\rt)
\ea
at late times. From this expression, we can see that the late-time complexity is independent of the magnetic field. The late-time rate vanishes in the purely magnetic black hole. However, this result will produce a puzzle in the limit of zero charges. For the purely magnetic static black hole, in the chargeless case $P\to 0$, it will reduce to a neutral black hole, and most of them capture the nonvanished late-time rate of the complexity. However, according to \eq{dIdt}, the late-time rate always vanishes, which implies that the chargeless limit also vanish. Therefore, in order to obtain an expected feature of the late-time rate at the zero charge limit, we need to add some extra boundary terms related to the electromagnetic field such that the late-time rate is sensitive to the magnetic charges.

\subsection{CA conjecture with some additional boundary term}\label{Sec32}
\subsubsection{Maxwell boundary term}
In order to obtain an expected feature of the complexity, with similar consideration of \cite{Goto:2018iay}, we also modify the action with the addition of the Maxwell boundary term. According to the equation of motion for the electromagnetic field, the Maxwell boundary term can be chosen as
\ba\label{mQbd}
I_{\m\text{Q}}=\g\int_{\pd M}\bm{G}\wedge \bm{A}\,,
\ea
where $\g$ is a free parameter, which should be determined by demanding that the holographic complexity shares expected feature under the zero-charge limit. Then, the general total action is given by
\ba
I_\text{total}=I+I_{\m\text{Q}}\,.
\ea
Adding this boundary term will give different boundary conditions. If the electromagnetic field satisfies the equation of motion $d\bm{G}=0$, using the Stokes' theory, this boundary term is equivalent to
\ba\label{ImQM}
I_{\m\text{Q}}=\int_M \bm{L}_{\m\text{Q}}
\ea
with
\ba\label{LmQ}
\bm{L}_{\m\text{Q}}=\g\bm{G}\wedge \bm{F}\,.
\ea
Its variation can be written as
\ba\begin{aligned}\label{dLmQ}
\g^{-1}\d \bm{L}_{\m\text{Q}}&=\d \bm{G}\wedge \bm{F}+\bm{G}\wedge \d \bm{F}=d\lf(\d \bm{B}\wedge \bm{F}+\bm{G}\wedge \d \bm{A}\rt)\,.
\end{aligned}\ea
By setting $\d =\math{L}_\z $ for any vector field $\z$, \eq{dLmQ} becomes
\ba\begin{aligned}
\g^{-1}d(\z\cdot \bm{L}_{\m\text{Q}})&=d\lf(\math{L}_\z \bm{B}\wedge \bm{F}+\bm{G}\wedge \math{L}_\z \bm{A}\rt)=d\lf[(\z\cdot \bm{G})\wedge \bm{F}+\bm{G}\wedge(\z \cdot\bm{F})\rt]\,,
\end{aligned}\ea
which implies that there exists a Noether charge $(n-2)$-form $\bm{K}_{\m\text{Q}}$ such that
\ba
\g^{-1}\z\cdot \bm{L}_{\m\text{Q}}-(\z\cdot \bm{G})\wedge \bm{F}-\bm{G}\wedge(\z \cdot\bm{F})=d \bm{K}_{\m\text{Q}}\,.
\ea
By using \eq{LmQ}, one can easy verify that $d \bm{K}_{\m\text{Q}}=0$. Then, we have
\ba\label{LHF1}
\g^{-1}\z\cdot \bm{L}_{\m\text{Q}}=(\z\cdot \bm{G})\wedge \bm{F}+\bm{G}\wedge(\z \cdot\bm{F})\,,
\ea

If we set $\z$ to be the static Killing vector field $\x$, \eq{LHF1} can be expressed as
\ba\label{LQ}
\x\cdot \bm{L}_{\m\text{Q}}=\g d(\F \bm{G}-\Y \bm{F})\,.
\ea
where we have used $d\bm{G}=d\bm{F}=0$.

Next, we start to evaluate its contribution to the holographic complexity. According to \eq{ImQM}, evaluating this additional boundary term can be translated into a bulk integration. Then, with similar procedures in the former section, the change of this additional action can be obtained by
\ba\begin{aligned}
I_{\m\text{Q}}&=\int_{\d M_2} \bm{L}_{\m\text{Q}}-\int_{\d M_1} \bm{L}_{\m\text{Q}}=-\g\d t\lf[\F_\math{H} Q-\Y_\math{H} P\rt]_+^-\,.
\end{aligned}\ea
The total action within the WDW patch is given by
\ba\begin{aligned}
\frac{d I_\text{total}}{d t}&=\frac{d}{d t}(I+I_{\m\text{Q}})=\left[(1-\g)\F_\math{H}Q+\g \Y_\math{H} P\right]^-_+\,.
\end{aligned}\ea
Subsequently, the final result for the late-time complexity growth rate becomes
\ba\begin{aligned}\label{dcdabd}
\lim_{t\to\inf}\frac{d C_\text{A}}{d t}=\frac{1}{\p \hbar}\left[(1-\g)\F_\math{H}Q+\g \Y_\math{H} P\right]^-_+\,.
\end{aligned}\ea

\subsubsection{Scalar boundary term}
In this subsection, we consider the following boundary term for the source-free scalar field
\ba
I_\f= \frac{\g_\f}{16\p}\int_{\pd M} \bm{Z}
\ea
with
\ba\begin{aligned}
\bm{Z}_{bcd}=\bm{\epsilon}_{abcd}\frac{\pd \math{L}_\f}{\pd \grad_a \f}\f\,.
\end{aligned}\ea

Similar with the Maxwell boundary term, this term modifies the character of the boundary condition of the scalar field. By using the Stokes' theory, it can be written as a bulk integration
\ba\begin{aligned}
I_\f=\frac{\g_\f}{16\p} \int_{M} d\bm{Z}\,.
\end{aligned}\ea

Next, we evaluate its contribution to the WDW patch. At the late time, the change of this contribution can be expressed as
\ba\begin{aligned}
I_\f&=\frac{\g_\f \d t}{16\p} \int_{\math{N}_2} \x\cdot d\bm{Z}-\frac{\g_\f \d t}{16\p} \int_{\math{N}_1} \x\cdot d\bm{Z}\\
&=\frac{\g_\f \d t}{16\p} \int_{\math{N}_1} d(\x\cdot\bm{Z})-\frac{\g_\f \d t}{16\p} \int_{\math{N}_2} d(\x\cdot\bm{Z})\\
&=\frac{\g_\f \d t}{16\p} \int_{\math{C}_-}\x\cdot\bm{Z}-\frac{\g_\f \d t}{16\p} \int_{\math{C}_+}\x\cdot\bm{Z}\,.
\end{aligned}\ea
Since $\x$ vanishes and $\bm{Z}$ is well-defined on the bifurcation surface, it is clear that $\x\cdot\bm{Z}$ also vanishes on the horizon, i.e., $I_\f=0$. Hence, adding this scalar boundary term does not change the complexity growth rate at late times for the multiple-horizon black hole.

\section{Dyonic RN black hole in $f(R)$ gravity}\label{Sec4}
In this section, we first apply our late-time result to a dyonic RN-AdS black hole for Maxwell-$f(R)$ gravity, where the bulk action is given by
\ba\label{actionfr}
I_\text{bulk}=\frac{1}{16\p}\int_M \bm{\epsilon}\lf[f(R)-F_{ab}F^{ab}\rt]
\ea
with the Ricci Scalar $R$. Then, we have
\ba
\bm{G}=\frac{1}{4\p} *\bm{F}\,\ \ \text{and}\ \ Q=\frac{1}{4\p}\int_{\math{C}_\inf}*\bm{F}\,.
\ea

By using \eq{mQbd}, the Maxwell boundary term can be expressed as
\ba
I_{\m\text{Q}}=\frac{\g}{4\p}\int_{\pd M}\bm{A}\wedge *\bm{F}\,.
\ea
According to \eq{actionfr}, the equation of motion can be expressed as
\ba\begin{aligned}\label{eomfr}
f'(R) R_{ab}-\frac{f(R)}{2}g_{ab}-(\grad_a \grad_b-g_{ab}\grad^c\grad_c)f'(R)=\frac{1}{2}T_{ab},\nn
\end{aligned}\\\ea
with the stress tensor of the electromagnetic field
\ba\label{Tabfr}
T_{ab}=4F_{ac}F_b{}^c-g_{ab}F_{cd}F^{cd}\,.
\ea

Next, we consider s special case, in which there exists an $R_0$ such that
\ba
f(R_0)=\frac{R_0}{2}f'(R_0)\,.
\ea
For the special case $R=R_0$, the equation of motion \eq{eomfr} becomes
\ba
R_{ab}-\frac{R_0}{4}g_{ab}=\frac{1}{2f'(R_0)}T_{ab}\,,
\ea
which implies that the dynoic Reissner-Nordstrom-AdS black hole with $L^2=-12/R_0$ is the solution of this theory. Its line element
can be described by the following metric,
\ba
ds^2=-b(r)dt^2+\frac{dr^2}{b(r)}+r^2(d\q^2+\sin^2\q d\f^2)
\ea
with the blackening factor
\ba
b(r)=\frac{r^2}{L^2}+1-\frac{2M}{r}+\frac{q_e^2+q_m^2}{r^2}\,.
\ea
The electromagnetic field can be written as
\ba\begin{aligned}
\bm{A}&=\sqrt{f'(R_0)}\lf[q_m(1-\cos\q)d\f-\frac{q_e}{r}dt\rt]\,,\\
\bm{F}&=\sqrt{f'(R_0)}\lf[-\frac{q_e}{r^2}dt\wedge dr+q_m \sin\q d\q\wedge d\f\rt]\,.
\end{aligned}\ea
And the Arnowitt-Deser-Misner mass is given by \cite{DKH}
\ba
M_F=f'(R_0)M\,.
\ea
By using these expressions, one can also obtain
\ba\begin{aligned}
\bm{G}&=\frac{\sqrt{f'(R_0)}}{4\p}\lf(\frac{q_m}{r^2}dt\wedge dr+q_e \sin\q d\q\wedge d\f\rt)\,,\\
\bm{B}&=\frac{\sqrt{f'(R_0)}}{4\p}\lf[q_e(1-\cos\q)d\f+\frac{q_m}{r}dt\rt]\,.
\end{aligned}\ea
Then, we have
\ba\begin{aligned}
Q&=\sqrt{f'(R_0)}q_e\,,\ \ \ \ \ \ \ \F_{\math{H}_\pm}=\sqrt{f'(R_0)}\frac{q_e}{r_\pm}\,,\\
P&=4\p \sqrt{f'(R_0)} q_m\,,\ \ \ \Y_{\math{H}_\pm}=\frac{\sqrt{f'(R_0)}}{4\p}\frac{q_m}{r_\pm}\,.
\end{aligned}\ea
The late-time CA complexity growth rate with the Maxwell boundary term can be expressed as
\ba\begin{aligned}
\lim_{t\to\inf}\frac{d C_\text{A}}{d t}&=\left.f'(R_0)\frac{(1-\g)q_e^2+\g q_m^2}{\p \hbar r}\right|^{r_-}_{r_+}\\
&=\frac{2M_F}{\p\hbar}\frac{(1-\g)q_e^2+\g q_m^2}{q_e^2+q_m^2}\,.
\end{aligned}\ea
When we consider the Einstein gravity $f(R)=R+6/L^2$, we can see that this result is same as that obtained by \cite{Goto:2018iay}. Then, in order to obtain the expected feature under the zero-charge limit, we need to set the coefficient $\g$ to satisfy
\ba
\g=\frac{q_m^2}{q_m^2-q_e^2}\,.
\ea

\section{Charged dilaton black hole}\label{Sec5}
In this section, we consider the charged dilation black hole for the Einstein gravity coupled to a dilaton field as well as a Maxwell field,
\ba\begin{aligned}
I_\text{bulk}&=\frac{1}{16\p}\int_{M} \bm{\epsilon}\lf[R-2(\grad \f)^2-V(\f)-e^{-2\a\f}\math{F}\rt]\,,\nn
\end{aligned}\\\ea
where the dilaton potential $V(\f)$ is given by \cite{GWGubbons}
\ba\begin{aligned}
V(\f)&=-\frac{2}{(1-\a^2)^2L^2}\lf[\a^2(3\a^2-1)e^{-2\f/\a}+(3-\a^2)e^{2\a\f}+8\a^2e^{(\a-1/\a)\f}\rt]\,.
\end{aligned}\ea
By using \eq{mQbd}, the Maxwell boundary term and scalar boundary term are expressed as
\ba\begin{aligned}\label{IIf}
I_{\m\text{Q}}=\frac{\g}{4\p}\int_{\pd M}e^{-2\a\f}\bm{A}\wedge \star\bm{F}\,,\ \ \ \ I_\f=\frac{\g_\f}{4\p}\int_{\pd M} \f \star d\f\,.
\end{aligned}\ea
Then, we consider the electrically charged dilaton black hole, which is given by \cite{CJGao}
\ba\begin{aligned}\label{dsf}
ds^2=-b(r)dt^2+\frac{dr^2}{b(r)}+U^2(r) (d\q^2+\sin^2\q d\f^2)\,,
\end{aligned}\ea
with
\ba\begin{aligned}
b(r)&=\lf(1-\frac{2M}{r}\rt)\lf(1-\frac{c}{r}\rt)^{\frac{1-\a^2}{1+\a^2}}+\frac{U^2(r)}{L^2}\,,\\
U^2(r)&=r^2\lf(1-\frac{c}{r}\rt)^{\frac{2\a^2}{1+\a^2}}\,.
\end{aligned}\ea
The electromagnetic field and dilaton field are written as
\ba\begin{aligned}\label{AF}
\bm{A}=-\frac{q_e}{r} dt\,,\ \ \bm{F}=-\frac{q_e}{r^2}dt\wedge dr\,,\ \ \f=\frac{1}{\a} \ln \lf(\frac{U(r)}{r}\rt)\,
\end{aligned}\ea
with
\ba
q_e^2=\frac{2M c}{1+\a^2}\,.
\ea
By using the relation
\ba\begin{aligned}
\bm{G}&=\frac{q_e \sin\q}{4\p} d\q\wedge d\f\,
\end{aligned}\ea
and \eq{AF}, we can find $Q=q_e$ and $\F=q_e/r$. In this paper, we only consider the black hole with multiple horizons, i.e., the case with $\a^2<1/3$. Then, the late-time complexity growth rate \eq{dcdabd} becomes
\ba\begin{aligned}\label{dcdabd}
\lim_{t\to\inf}\frac{d C_\text{A}}{d t}=\left.\frac{(1-\g)q_e^2}{\p \hbar r}\right|^{r_-}_{r_+}\,.
\end{aligned}\ea
Next, we consider the scalar boundary term. By using \eq{dsf}, we can obtain
\ba
\bm{Z}=-4b(r) U^2(r) \f \pd_r\f \sin\q dt\wedge d\q \wedge d\f\,.
\ea
And the scalar boundary term can be written as
\ba\begin{aligned}
I_\f&=\frac{\g_\f}{16\p} \int_{\math{C}_+}\x\cdot\bm{Z}-\frac{\g_\f}{16\p} \int_{\math{C}_-}\x\cdot\bm{Z}\\
&=\left.\frac{\g_\f}{4\p}U^2(r)b(r)\f\pd_r\f\right|^{r_-}_{r_+}\\
&=0\,.
\end{aligned}\ea

The neutral case can be obtained by setting $c\to 0$. Then, the late-time growth rate becomes
\ba\begin{aligned}\label{dcdabd}
\lim_{t\to\inf}\frac{d C_\text{A}}{d t}=\frac{2M}{\p \hbar}\frac{1-\g}{1-\a^2}\,.
\end{aligned}\ea
In order to obtain the expected feature of the zero-charge limit, we need set the coefficient $\g$ to satisfy $\g=\a^2$.

\section{Some static magnetic black holes in Einstein gravity}\label{Sec6}
In this section, we will first apply our late-time result \eq{dcdabd} to some explicit magnetic black holes in Einstein gravity and discuss which conditions can give an expected feature of the complexity at zero-charge limit. Then, we will generally discuss the proper condition for the static magnetic black holes in Einstein gravity coupled to a first-order electromagnetic field.

\subsection{Bardeen black hole}\label{Sec61}

In this subsection, we consider the Bardeen black hole for the nonlinear gauge theories. The bulk action can be written as
\ba
I_\text{bulk}=\frac{1}{16\p}\int_{M}\bm{\epsilon}\lf(R+\frac{6}{L^2}-h(\math{F}^{(2)})\rt)\,.
\ea
In \cite{Bardeen}, Bardeen first proposed a black hole solution being regular at $r=0$ where the standard black hole spacetime has a physical singularity.
In this subsection, we consider the AdS-Bardeen spacetime, which can be described by \cite{AyonBeato:2000zs,Fan:2016hvf}
\ba\label{ds2Bardeen}
ds^2=-b(r)dt^2+\frac{dr^2}{b(r)}+r^2(d\q^2+\sin^2\q d\f^2)
\ea
with
\ba
b(r)=\frac{r^2}{L^2}+1-\frac{2M r^2}{(r^2+q_m^2)^{3/2}}\,.
\ea
This spacetime is parameterised by the mass parameter $M$ and the magnetic charge $q_m$. It is not hard to verify that this spacetime
is a solution of the Einstein gravitational equation coupled to nonlinear electromagnetic field with
\ba\label{hBardeen}
h(\math{F})= \frac{12 M}{q_m^3}\lf(\frac{\sqrt{-q_m^2\math{F}^{(2)}/2}}{1+\sqrt{-q_m^2\math{F}^{(2)}/2}}\rt)^{5/2}\,.
\ea
For the AdS-Bardeen solution, the electromagnetic field is given by
\ba\begin{aligned}
\bm{A}=q_m(1-\cos\q)d\f\,,\ \ \ \bm{F}=q_m \sin\q d\q\wedge d\f\,,
\end{aligned}\ea
which gives
\ba
\math{F}^{(2)}=-F_{ab}F^{ab}=-\frac{2q_m^2}{r^4}\,.
\ea
From \eq{Hab}, we can obtain
\ba\begin{aligned}
\bm{G}&=\frac{15M q_mr^4}{8\p (q_m^2+r^2)^{7/2}}dt\wedge dr\,,\\
\bm{B}&=\frac{3M}{8\p q_m}\lf[1-\frac{r^5}{(q_m^2+r^2)^{5/2}}\rt]dt\,.
\end{aligned}\ea
According to these expressions, we can find the magnetic potential and charge,
\ba\begin{aligned}
\Y&=\frac{3M}{8\p q_m}\lf[1-\frac{r^5}{(q_m^2+r^2)^{5/2}}\rt]\,,\\
P&=4\p q_m\,.
\end{aligned}\ea
As a result, the late-time CA complexity growth rate with the Maxwell boundary term can be expressed as
\ba\begin{aligned}
\lim_{t\to\inf}\frac{d C_\text{A}}{d t}&=\left.\frac{3\g M}{2\p \hbar}\frac{r^5}{(q_m^2+r^2)^{5/2}}\right|^{r_+}_{r_-}\,,
\end{aligned}\ea
which becomes
\ba\begin{aligned}
\lim_{t\to\inf}\frac{d C_\text{A}}{d t}&=\frac{3\g M}{2\p \hbar}\,.
\end{aligned}\ea
under the zero-charge limit. In order to obtain the expected feature of the zero-charge limit, we need set the coefficient $\g=4/3$ such that
\ba\begin{aligned}
\lim_{t\to\inf}\frac{d C_\text{A}}{d t}&=\frac{2M}{\p \hbar}\,.
\end{aligned}\ea
under the limit $q_m\to 0$.

\subsection{Static magnetic black hole in Einstein-$\math{F}^{(2n)}$ gravity}\label{Sec62}
In this subsection, we consider the static magnetic solution for Einstein gravitational theory coupled a electrodynamics with the lagrangian $h(\math{F})=(-1)^n\math{F}^{(2n)}$. The equation of motion can be expressed as
\ba\begin{aligned}\label{eom}
R_{ab}-\frac{1}{2}Rg_{ab}-\frac{3}{L^2}g_{ab}&=\frac{1}{2} T_{ab}\,,\\
\grad^a H^{(2n-1)}_{ab}=0\,,
\end{aligned}\ea
with
\ba\begin{aligned}
T_{ab}&=(-1)^{n-1}4nH_{ac}^{(2n-1)}F_b{}^c-(-1)^n\math{F}^{(2n)}g_{ab}\,.
\end{aligned}\ea
As mentioned above, we next consider the geometry of the static purely magnetic black hole solution. Its not hard to verify that the spherically static solution can be written as
\ba\begin{aligned}\label{ds2D}
ds^2&=-b(r)dt^2+\frac{dr^2}{b(r)}+r^2(d\q^2+\sin^2\q d\f^2)\,,\\
\bm{A}&=q_m(1-\cos\q) d\f\,,
\end{aligned}\ea
with the blackening factor
\ba
b(r)=\frac{r^2}{L^2}+1-\frac{2M}{r}+\frac{q_m^{2n}}{(4n-3)r^{4n-2}}\,.
\ea
According to the solution \eq{ds2D}, we can further obtain
\ba\begin{aligned}
\bm{F}&= q_m \sin\q d\q\wedge d\f\,,\\
\bm{G}&= (-1)^{n-1}\frac{n}{4\p} \star \bm{H}^{(2n-1)}=\frac{n}{4\p}\frac{q_m^{2n-1}}{r^{4n-2}}dt\wedge dr\,,
\end{aligned}\ea
which implies
\ba
\bm{B}=\frac{n}{4(4n-3)\p}\frac{q_m^{2n-1}}{r^{4n-3}}dt\,.
\ea
And the magnetic potential and charge can be read off
\ba\begin{aligned}
\Y&=\frac{n}{4(4n-3)\p}\frac{q_m^{2n-1}}{r^{4n-3}}\,,\\
P&=4\p q_m\,.
\end{aligned}\ea
Using these expressions, the late-time CA complexity rate can be shown as
\ba
\lim_{t\to\inf}\frac{dC_A}{dt}=\lf.\frac{\g n}{(4n-3)\p \hbar}\frac{q_m^{2n}}{r^{4n-3}}\rt|^{r_+}_{r_-}\,.
\ea
At the chargeless limit $q_m\to0$, the action growth rate becomes
\ba
\lim_{t\to\inf}\frac{dC_A}{dt}=\frac{2\g n M}{\p \hbar}\,.
\ea
In order to obtain the expected feature of the zero-charge limit, we need to set the coefficient $\g$ to satisfy $\g=1/n$.

\subsection{A general discussion for the static magnetic black holes with first-order electrodynamics}\label{Sec63}
In the former subsections, we applied our late-time result \eq{dcdabd} to some explicit cases of the magnetic black hole in Einstein gravity and showed how to choose the boundary terms to make the complexity be well-defined in the zero-charge limit. From these case, we can see that the choice of the proportional constant is dependent on the explicit case of the electromagnetic theory as well as the spacetime background. In this subsection, we will generally study the static magnetic black hole in Einstein gravity coupled to a first-order electromagnetic field, where the bulk action is shown as
\ba
I_\text{bulk}=\frac{1}{16\p}\int_{M}\bm{\epsilon}\lf(R+\frac{6}{L^2}-h(\math{F})\rt)\,.
\ea
The equation of motion of the gravity part can be read off
\ba\begin{aligned}\label{eom6}
R_{ab}-\frac{1}{2}Rg_{ab}-\frac{3}{L^2}g_{ab}&=\frac{1}{2} T_{ab}
\end{aligned}\ea
with
\ba
T_{ab}&=H_{ac}F_b{}^c-g_{ab}h(\math{F})\,,
\ea
where $H_{ab}$ is defined in \eq{Hab} with
\ba
h_n=\frac{\pd h(\math{F})}{\pd \math{F}^{(2n)}}
\ea
Here we assume that $h(\math{F})$ only vanishes when the electromagnetic filed vanishes, i.e., $h(\math{F})=0$ iff $\math{F}=0$. Without loss of generality, we next consider the geometry of the static regular magnetic black hole with the metric and electromagnetic filed ansatz
\ba\begin{aligned}\label{dsA}
ds^2&=-f(r)dt^2+\frac{dr^2}{f(r)}+r^2\lf(d\q^2+\sin^2\q d\f^2\rt)\,,\\
\bm{A}&=q_m(1-\cos\q) d\f\,,
\end{aligned}\ea
with the blackening factor
\ba\label{fr6}
f(r)=1+\frac{r^2}{L^2}-\frac{2m(r)}{r}\,.
\ea
At the zero-charge limit, this solution should reduce to the SAdS solution, i.e., $m(r)=M$ when $q_m\to 0$. Moreover, we also assume that this solution shares the similar behavior with the SAdS black hole at the asymptotic infinity, i.e., $m(r)=M$ when $r\to \inf$.
Using this solution ansatz, one can further obtain
\ba\begin{aligned}\label{HF6}
\bm{F}&= q_m \sin\q d\q\wedge d\f\,,\\
\bm{H}&=q(r)\sin\q d\q\wedge d\f\,,
\end{aligned}\ea
with
\ba
q(r)=4\sum_n (-1)^{n-1}\frac{n h_nq_m^{2n-1}}{r^{4(n-1)}}\,.
\ea
According to the equation of motion \eq{eom6}, we find that there are only two independent equations, i.e.,
\ba\begin{aligned}\label{meom}
&4 m'(r)-h(\math{F}) r^2=0\,,\\
&h(\math{F})r^4-2r^3m''(r)-q_m q(r)=0\,,
\end{aligned}\ea
which give
\ba\label{qr}
q(r)=\frac{2r^2[2m'(r)-r m''(r)]}{q_m}\,.
\ea
Combining with \eq{HF6}, one can further find
\ba
\bm{G}=\frac{q(r)}{16\p r^2} dt\wedge dr=\frac{2m'(r)-r m''(r)}{8\p q_m}dt\wedge dr\,,
\ea
which implies
\ba
\bm{B}=\frac{rm'(r)-3m(r)}{8\p q_m} dt\,.
\ea
With these in mind, we can obtain
\ba
P=4\p q_m\,,\ \ \Y=\frac{rm'(r)-3m(r)}{8\p q_m}\,.
\ea
Then, the late-time CA complexity growth rate is given by
\ba\label{CA6}
\lim_{t\to\inf}\frac{dC_A}{dt}=\left.\frac{\g\lf[3m(r)-h(\math{F})r^3/4\rt]}{2\p \hbar}\right|_{r_-}^{r_+}\,.
\ea
Next, we consider the zero-charge limit of the late-time rate. From the solution ansatz, we have
\ba\label{F2n6}
\math{F}^{(2n)}=(-1)^{n} \frac{2q_e^{2n}}{r^{4n}}\,,
\ea
which gives $\math{F}(r_+)=0$ under the zero-charge limit $q_m\to0$. This implies $h\lf(\math{F}(r_+)\rt)=0$ under $q_m\to0$. Moreover, from the blackening factor \eq{fr6}, we have
\ba\label{mrm}
m(r_-)=\frac{r_-}{2}+\frac{r_-^3}{2L^2}\to 0
\ea
under the zero charge limit. Then, the late-time complexity growth rate \eq{CA6} becomes
\ba\label{CA62}
\lim_{t\to\inf}\frac{dC_A}{dt}=\frac{3\g M}{2\p \hbar}+\lim_{q_e\to 0}\lf[\frac{\g h(\math{F}) r^3}{8\p \hbar}\rt]_{r_-}
\ea
at the limit $q_e\to0$. The key point to obtain \eq{CA62} is to find the behaviour of $h\lf(\math{F}(r_-)\rt)$ under the zero-charge limit. From \eq{F2n6}, we can see that $h(\math{F})$ can be expressed as a function of $x=q_m/r^2$, i.e., $h(\math{F})=h(x)$. According to \eq{meom}, the mass function can be expressed as
\ba\begin{aligned}
m(r)&=M+\frac{1}{4}\int_{0}^{r}drr^2h(x)\\
&=M+\frac{q^{3/2}_m}{4}\int_{0}^{x}dxx^{-5/2}h(x)\\
&=M-q^{3/2}_m \tilde{m}(x)\,,
\end{aligned}\ea
where we denote
\ba\label{mx}
\tilde{m}(x)=-\frac{1}{4}\int_{0}^{x}dxx^{-5/2}h(x)\,.
\ea
The asymptotic condition $m(r\to\inf)=M$ implies $\tilde{m}(x\to0)=0$. Combing \eq{mx} with the limit \eq{mrm}, we have
\ba\label{mxm}
\tilde{m}(x_-)\simeq \frac{M}{q_m^{3/2}}\to \inf
\ea
when $q_m\to 0$. Since $h(x)$ is a smooth function, this equation implies $x_-\to \inf$ under the zero-charge limit. Then, there are two cases we should consider, that is, $h\lf(\math{F}\rt)$ being convergent or divergent.

(a) If $h(\math{F})$ is a convergent function, we have $h(x_-) r_-^3 \to 0$ under the zero charge limit. The late-time growth becomes
\ba\label{CA63}
\lim_{t\to\inf}\frac{dC_A}{dt}=\frac{3\g M}{2\p \hbar}
\ea
under the zero-charge limit. In order to obtain the expected feature of this limit, we need to set $\g=4/3$. This implies that the choice of $\g$ is independent on the explicit expression of $h(\math{F})$ if $h(\math{F})$ is convergent. We can see that the Bardeen black hole is exactly this situation.

(b) Next, we consider the case where $h(\math{F})$ is a divergent function. Eq. \eq{mxm} implies that we can only consider the asymptotic behavior of $h(x)$. In this paper, we suppose that $h(\math{F})$ has the asymptotic behavior
\ba\label{max}
h(x)\simeq a_0 x^{2\n}=\frac{a_0q_m^{2\n}}{r^{4\n}}\,.
\ea
According to the equation of motion \eq{meom}, one can further obtain
\ba
m(r)\simeq M-\frac{a_0q_m^{2\n}}{4(4\n -3)r^{4\n-3}}\,,
\ea
which implies
\ba
h(\math{F})r^3_-=\frac{a_0q_m^{2\n}}{r^{4\n-3}}\simeq 4(4\n-3)M
\ea
at $q_m\to 0$. Then, the zero-charge limit of \eq{CA62} gives
\ba
\lim_{t\to\inf}\frac{dC_A}{dt}=\frac{2\g \n M}{\p \hbar}\,.
\ea
In order to obtain the expected feature of the zero-charge limit, we need to set the coefficient $\g$ to satisfy $\g=1/\n$. The case of $h(\math{F})=\math{F}^{(2n)}$ in the last subsection is actually this situation with $\n=n$.

\section{Conclusion}\label{Sec7}
Motivated by \cite{Goto:2018iay} where the vanishing of the late-time CA complexity rate in purely magnetic $dyonic$ RN-AdS black hole was found and a remedy was proposed, in this paper, we evaluated the original CA holographic complexity in a static multiple-horizon black hole for a gravitational theory coupled to a first-order source-free electrodynamics. We showed that the vanishing feature of the late-time rate in the purely magnetic black hole is universal for the original CA conjecture. But this result does not agree with the general expectation \eq{TS} of the quantum system, and it also has an unexpected feature in the zero-charge limit. However, these failures could be alleviated when we modified the action with an additional term (Maxwell boundary term) within the WDW patch. By  Iyer-Wald formalism, we generally showed the late-time complexity growth rate after adding Maxwell boundary term. We also found that the scalar boundary term does not change the late-time rate for a multiple-horizon black hole with source-free electrodynamics. Moreover, there exists a dimensionless parameter $\g$ which is needed to be chosen by demanding the zero-charge limit satisfies. To be specific, we applied our result to the $dyonic$ RN black hole in  $f(R)$ gravity, charged dilation black hole, Bardeen black hole, and the static magnetic black hole in Einstein gravity coupled a electromagnetic field with $h(\math{F})=(-1)^n\math{F}^{2n}$. We found that the proper proportional parameter $\g$ is dependent on specific gravitational theory and the spacetime background. Finally, we investigated the static magnetic black hole for the Einstein gravity coupled to a general first-order electromagnetic field and found the relationship between the proper proportional constant and the Lagrangian function $h(\math{F})$ of the electromagnetic field. If $h(\math{F})$ is a convergent function, we need to choose $\g=4/3$; if $h(\math{F})$ is a divergent function with the asymptotic behavior \eq{max}, we need to choose $\g=\n^{-1}$.

\section*{acknowledgment}
 This research is supported by NSFC Grants No. 11775022 and 11375026.

\end{document}